\documentclass[a4paper,twoside,12pt]{article}
\input{obsjkt.sty}

\newcommand{\Msun}{\ensuremath{~{\rm M}_\odot}}                   % Solar mass symbol
\newcommand{\Rsun}{\ensuremath{~{\rm R}_\odot}}                   % Solar radius symbol
\newcommand{\Teff}{\ensuremath{T_{\rm eff}}}                      % Effective temperature symbol
\newcommand{\EBV}{\ensuremath{E(B\!-\!V)}}                        % E(B-V) symbol
\newcommand{\degr}{\ensuremath{^\circ}}                           % Degree symbol
\renewcommand{\kms}{~km~s$^{-1}$}                                 % km/s symbol
\newcommand{\as}{\ensuremath{^{\prime\prime}}}                    % Arcsecond symbol
\newcommand{\kepler}{\textit{Kepler}}

\newcommand{\gaia}{\textit{Gaia}}
\newcommand{\targ}{V1765~Cyg}
\newcommand{\targfull}{V1765~Cygni}

\newcommand{\Msunnom}{\hbox{$\mathcal{M}^{\rm N}_\odot$}}
\newcommand{\Rsunnom}{\hbox{$\mathcal{R}^{\rm N}_\odot$}}
\newcommand{\Lsunnom}{\hbox{$\mathcal{L}^{\rm N}_\odot$}}

\usepackage{rotating}

\begin{document} %%%%%%%%%%%%%%%%%%%%%%%%%%%%%%%%%%%%%%%%%%%%%%%%%%%%%%%%%%%%%%%%%%%%%%%%%%%%%%%%%%%%%%%%%%%%%%%%%%%%%%%%%%%%%%%%%%%%%%%%%%%%%%%%%%%%
%%%%%%%%%%%%%%%%%%%%%%%%%%%%%%%%%%%%%%%%%%%%%%%%%%%%%%%%%%%%%%%%%%%%%%%%%%%%%%%%%%%%%%%%%%%%%%%%%%%%%%%%%%%%%%%%%%%%%%%%%%%%%%%%%%%%%%%%%%%%%%%%%%%%%

\OBSheader{Rediscussion of eclipsing binaries: \targ}{J.\ Southworth}{2023 October}

\OBStitle{Rediscussion of eclipsing binaries. Paper XV. \\ The B-type supergiant system V1765 Cygni}

\OBSauth{John Southworth}

\OBSinstone{Astrophysics Group, Keele University, Staffordshire, ST5 5BG, UK}

%\OBSabstract{Abs abs abs abs abs abs abs abs abs abs abs abs abs abs abs abs abs abs abs abs abs abs abs abs abs abs abs abs abs abs abs abs abs abs abs abs abs abs abs abs abs abs abs abs abs abs abs abs abs abs abs abs abs abs abs abs abs abs abs abs abs abs abs abs abs abs abs abs abs abs abs abs abs abs abs abs abs abs abs abs abs abs abs abs abs abs abs abs abs abs abs abs abs abs abs abs abs abs abs abs abs abs abs abs abs abs abs abs abs abs abs abs abs abs abs abs abs abs abs abs abs abs abs abs abs abs abs abs abs abs abs abs abs abs abs abs abs abs abs abs abs abs abs abs abs abs abs abs abs abs abs abs abs abs abs abs abs abs abs abs abs abs abs abs abs abs abs abs abs abs abs abs abs abs abs abs abs abs abs abs abs abs abs abs abs abs abs abs abs abs abs abs abs abs abs abs abs abs abs abs abs abs abs abs abs abs abs abs abs abs abs abs abs abs abs abs abs abs abs abs abs abs abs abs abs abs abs abs abs abs abs abs abs abs abs abs abs abs abs abs abs abs abs.}

\OBSabstract{\targ\ is a detached eclipsing binary containing a B0.5 supergiant and a B1 main-sequence star, with an orbital period of 13.37~d and an eccentricity of 0.315. The system shows apsidal motion and the supergiant exhibits strong stochastic variability. \targ\ was observed by the Transiting Exoplanet Survey Satellite over four sectors. We analyse these data to obtain the first determinate light curve model for the system. To this we add published spectroscopic orbits to infer masses of $23 \pm 2$ and $11.9 \pm 0.7$\Msun, and radii of $20.6 \pm 0.8$ and $6.2 \pm 0.3$\Rsun. These properties are in good agreement with theoretical predictions for a solar chemical composition and an age around 7~Myr. We also present two epochs of blue-optical spectroscopy that confirm the luminosity classification of the primary star and appear to show absorption lines from the secondary star. Extensive spectroscopy and further analysis of the system is recommended.}

%%%%%%%%%%%%%%%%%%%%%%%%%%%%%%%%%%%%%%%%%%%%%%%%%%%%%%%%%%%%%%%%%%%%%%%%%%%%%%%%%%%%%%%%%%%%%%%%%%%%%%%%%%%%%%%%%%%%%%%%%%%%%%%%%%%%%%%%%%%%%%%%%%%%%

\section*{Introduction}

Detached eclipsing binaries (dEBs) are a vital source of empirical measurements of the properties of stars \cite{Andersen91aarv,Torres++10aarv,Me15aspc}. Such measurements typically show a good agreement with theoretical predictions except for stars of very low or very high mass. At lower masses, M-dwarfs are know to show a \emph{radius discrepancy} which remains unsolved \cite{Hoxie73aa,Lacy77apjs,Torres13an}. At higher masses, there is a \emph{mass discrepancy} whereby stellar masses inferred from the stars' positions in the Hertzsprung-Russell diagram are systematically greater than those measured directly from orbital motion in binary systems \cite{Herrero+92aa}. Tkachenko et al.\ \cite{Tkachenko+20aa} have investigated this in detail using dEBs and concluded that it is stronger at lower surface gravities, and is partially caused by overestimation of the effective temperatures (\Teff s) of massive stars from their optical spectra.

Massive stars are typically found in multiple systems \cite{Sana+14apjs,Kobulnicky+14apjs} and most also show brightness variability due to a range of phenomena. Massive stars in dEBs have been found to show intrinsic variations due to stochastic low-frequency (SLF) variability \cite{Bowman+19aa,MeBowman22mn} and $\beta$\,Cephei pulsations \cite{Me+20mn,LeeHong21aj,Me++21mn}. At lower masses, A- and F-stars in dEBs can show variability due to $\delta$\,Scuti \cite{Kahraman+17mn,Chen+22apjs,Me21obs6,Me++23mn} and $\gamma$\,Doradus pulsations \cite{Debosscher+13aa,Lee16apj,MeVanreeth22mn}. In all cases the pulsations can be perturbed or excited by tidal effects in close binary systems \cite{Kurtz+20mn,Handler+20natas,Me21obs6,MeVanreeth22mn}.

In this work we present the first analysis of extensive space-based photometry for the bright B-type supergiant system \targ\ (Table~\ref{tab:info}), which has a long observational history. The new photometric data also exhibit a strong signature of SLF variability. See ref.~\cite{Me20obs} for a detailed description of our project and ref.~\cite{Me21univ} for a review of the impact of space-based photometry on binary star science.

%%%%%%%%%%%%%%%%%%%%%%%%%%%%%%%%%%%%%%%%%%%%%%%%%%%%%%%%%%%%%%%%%%%%%%%%%%%%%%%%%%%%%%%%%%%%%%%%%%%%%%%%%%%%%%%%%%%%%%%%%%%%%%%%%%%%%%%%%%%%%%%%%%%%%

\section*{\targfull}

\begin{table}[t]
\caption{\em Basic information on \targ. \label{tab:info}}
\centering
\begin{tabular}{lll}
{\em Property}                            & {\em Value}                 & {\em Reference}                   \\[3pt]
Right ascension (J2000)                   & 19:48:50.60                 & \cite{Gaia21aa}                   \\
Declination (J2000)                       & +33:26:14.2                 & \cite{Gaia21aa}                   \\
Bright Star Catalogue                     & HR 7551                     & \cite{HoffleitJaschek91}          \\
Henry Draper designation                  & HD 187459                   & \cite{CannonPickering23anhar}     \\
% \textit{Hipparcos} designation          & HIP 97485                   & \cite{Hipparcos97}                \\
% \textit{Tycho} designation              & TYC 2673-65-1               & \cite{Hog+00aa}                   \\
\textit{Gaia} DR3 designation             & 2034968875123889536         & \cite{Gaia21aa}                   \\
\textit{Gaia} DR3 parallax                & $0.6895 \pm 0.0250$ mas     & \cite{Gaia21aa}                   \\          % d = 120.55 +/- 0.52 pc
TESS\ Input Catalog designation           & TIC 59632148                & \cite{Stassun+19aj}               \\
$B$ magnitude                             & $6.578 \pm 0.014$           & \cite{Hog+00aa}                   \\          % \cite{Henden+15aas} for APASS
$V$ magnitude                             & $6.463 \pm 0.010$           & \cite{Hog+00aa}                   \\          % \cite{Hog+00aa} for Tycho
$J$ magnitude                             & $6.027 \pm 0.019$           & \cite{Cutri+03book}               \\
$H$ magnitude                             & $6.034 \pm 0.018$           & \cite{Cutri+03book}               \\
$K_s$ magnitude                           & $6.030 \pm 0.020$           & \cite{Cutri+03book}               \\
Spectral type                             & B0.5\,Ib + B1\,V            & \cite{MorganRoman50apj}, This work\\[3pt]
\end{tabular}
\end{table}

\targ\ was announced as a spectroscopic binary by Plaskett \& Pearce \cite{PlaskettPearce31pdao}. The primary component (hereafter star~A) is a B0.5 supergiant and is the source of the observed SLF variability. The secondary component (star~B) is of similar \Teff\ but is much smaller than the supergiant component. For clarity, star~A is eclipsed (i.e.\ is at superior conjunction) by star~B at primary eclipse.

Mayer \& Chochol \cite{MayerChochol81pasp} discovered the eclipses and also commented on the presence of ``irregular fluctuations in the range of about 0.06~mag''. They also obtained radial velocities (RVs) of star~A and asserted the presence of apsidal motion. However, they did not attempt a solution of the light curve. Percy \& Welch \cite{PercyWelch83pasp} confirmed the presence of ``pronounced intrinsic variability''.

The spectral type of the much brighter component of the system has been given as either B0.5\,Ib \cite{MorganRoman50apj,Morgan++55apjs,Hiltner56apjs} or B0.5\,II \cite{Hiltner51apj,Lesh68apjs}. A classification of B0.5\,Ib + B2\,V was given by Hill \& Fisher \cite{HillFisher84aa} (hereafter HF84).

HF84 presented the most detailed analysis of the system thus far, based on photographic spectra subsequently converted to electronic format for analysis. They found star~B to show up reasonably well in He~I lines and determined RVs via cross-correlation. The plotted cross-correlation functions (HF84's fig.~1) show that the two components are never resolved, so the RVs were obtained by fitting double overlapping Gaussian functions. Apsidal motion was detected but at a level below that required for confirmation. HF84 presented mass and radius estimates for both components but relied on a calibration of radius versus spectral type for star~A as the solution of the light curves from Mayer \& Chochol \cite{MayerChochol81pasp} was indeterminate.

Mayer et al.\ \cite{Mayer+91baicz2} obtained new photometry and spectroscopy and estimated the masses and radii of the components, giving values similar to those found by HF84. However, they preferred an earlier spectral type for star~B of of B1\,V or even B0\,V. Raja \cite{Raja94aa} presented further photographic spectroscopy in which they were not able to find a sign of star~B, but found an apsidal motion of $7.3 \times 10^{-4}$~deg~d$^{-1}$, in agreement with previous results.

Popper \cite{Popper93pasp} presented a small number of high-quality spectra of \targ. He obtained a much larger rotational velocity of $\sim$200\kms\ for star~A, versus the value of $135 \pm 10$\kms\ measured by HF84. He was also not able to find clear evidence of spectral lines of star~B despite being sensitive to much smaller lines than expected based on the light ratio of the system inferred by HF84. He concluded that the system was unfavourable for further analysis due to the difficulties it poses for both spectroscopy and photometry.

Percy \& Khaja \cite{PercyKhaja95jrasc} presented further photometry of the system from which they measured the eclipse depths and found a possible slow increase in brightness. Since then, \targ\ has mostly been left well alone save for appearances in large sky surveys. The MASCARA cameras \cite{Talens+17aa} have observed \targ\ extensively and obtained 12\,057 photometry measurements of the system \cite{Burggraaff+18aa} which show the eclipses and intrinsic variability.

In the current work we use extensive new light curves to investigate the photometric properties of the system, infer its physical properties, examine two new spectra, and draw attention to the similarity between \targ\ and V380~Cyg. We conclude with a discussion on the future prospects for analysis of this important but challenging binary system.

% /home/jkt/papers/writtenup/1995JRASC..89...91P.ps.gz  PercyKhaja95jrasc
% /home/jkt/papers/writtenup/1994A+A...284...82R.ps.gz  Raja94aa
% /home/jkt/papers/writtenup/1993PASP..105..721P.ps.gz  Popper93pasp
% /home/jkt/papers/writtenup/1991BAICz..42..230M.ps.gz  Mayer+91baicz2
% /home/jkt/papers/writtenup/1984A+A...139..123H.ps.gz  HillFisher84aa
% /home/jkt/papers/notprinted/1983PASP...95..491P.pdf   PercyWelch83pasp 
% /home/jkt/papers/notprinted/1981PASP...93..608M.pdf   MayerChochol81pasp

%%%%%%%%%%%%%%%%%%%%%%%%%%%%%%%%%%%%%%%%%%%%%%%%%%%%%%%%%%%%%%%%%%%%%%%%%%%%%%%%%%%%%%%%%%%%%%%%%%%%%%%%%%%%%%%%%%%%%%%%%%%%%%%%%%%%%%%%%%%%%%%%%%%%%

\section*{Photometric observations}

\begin{figure}[t] \centering \includegraphics[width=\textwidth]{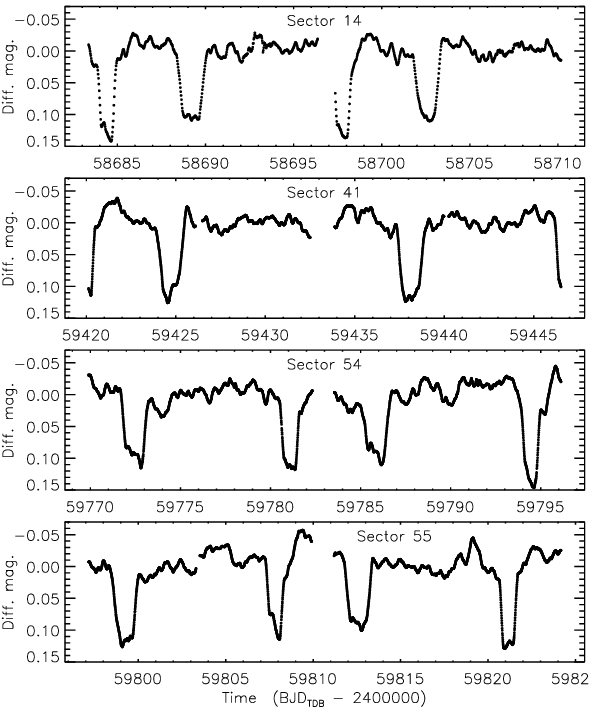} \\
\caption{\label{fig:time} TESS\ short-cadence SAP photometry of \targ. The flux 
measurements have been converted to magnitude units then rectified to zero magnitude 
by subtraction of the median. The individual sectors are labelled.} \end{figure}

% Sector 14 (2019-Jul-18 to 2019-Aug-15, in cycle 2): observed in camera 1.
% Sector 41 (2021-Jul-23 to 2021-Aug-20, in cycle 4): observed in camera 1.
% Sector 54 (2022-Jul-09 to 2022-Aug-05, in cycle 4): observed in camera 2.
% Sector 55 (2022-Aug-05 to 2022-Sep-01, in cycle 4): observed in camera 3.

\targ\ has been observed four times by the NASA Transiting Exoplanet Survey Satellite \cite{Ricker+15jatis} (TESS). The data from sector 14 (2019/07/18 to 2019/08/15) were obtained at a cadence of 1800~s, and the data from sector 41 (2021/07/23 to 2021/08/23) and sectors 54 and 55 (2022/07/09 to 2022/09/01) had an observing cadence of 600~s. We used the {\sc lightkurve} package \cite{Lightkurve18} to download the data and reject points flagged as bad. We adopted the simple aperture photometry (SAP) data \cite{Jenkins+16spie} for consistency with previous papers in this series.

We converted the data to differential magnitude and subtracted the median magnitude for further analysis. The numbers of datapoints are 1237, 3505, 3571 and 3645, for sectors 14, 41, 54 and 55, respectively. Data with a 120~s cadence are available for all but the first sector, but were not used in our analysis because the system does not vary on a timescale fast enough to require the higher sampling rate.  

The data are shown in Fig.~\ref{fig:time}. The light curves clearly show the existence of annular primary eclipses, total secondary eclipses, orbital eccentricity and SLF variability. 

%We queried the \gaia\ DR3 database\footnote{\texttt{https://vizier.cds.unistra.fr/viz-bin/VizieR-3?-source=I/355/gaiadr3}} for objects within 2~arcmin of \targ. A total of 108 were found, all of which are fainter than \targ\ by at least 7.2~mag in the \gaia\ $G$ band. We deduce that the amount of light contaminating the TESS aperture for this dEB is negligible.

%%%%%%%%%%%%%%%%%%%%%%%%%%%%%%%%%%%%%%%%%%%%%%%%%%%%%%%%%%%%%%%%%%%%%%%%%%%%%%%%%%%%%%%%%%%%%%%%%%%%%%%%%%%%%%%%%%%%%%%%%%%%%%%%%%%%%%%%%%%%%%%%%%%%%

\section*{Spectroscopic observations}

We obtained spectroscopy of \targ\ on the nights of 2023/07/02 and 2023/07/04 in order to investigate the suitability of the system for detailed analysis. Our observing run lasted 7 nights and covered only half of the orbit of \targ, so we made no attempt to obtain sufficient data for measuring the spectroscopic orbits of the stars. Instead we obtained two spectra on the first night at a time when the two stars had approximately the same RV, and three spectra on the second night when the stars were close to their largest RV separation. 

We used the Isaac Newton Telescope (INT), Intermediate Dispersion Spectrograph (IDS) with the 235~mm camera, the EEV10 CCD, the H2400B grating, a central wavelength of 420~nm, and a 1\as\ slit. This gave spectra with a reciprocal dispersion of 0.24~\AA~mm$^{-1}$, a resolution of 0.5~\AA\ as measured from the Cu+Ar lamps used for wavelength calibration, and a spectral coverage of 410--465~nm. The data were reduced using a pipeline currently under construction (see ref.~\cite{Me22obs6}).

The spectra from each night were taken at the same time so were summed to give two overall spectra. The total exposure time was 240~s on the first night and 720~s on the second night, with the latter being significantly longer to compensate for the presence of moderate cloud.

%%%%%%%%%%%%%%%%%%%%%%%%%%%%%%%%%%%%%%%%%%%%%%%%%%%%%%%%%%%%%%%%%%%%%%%%%%%%%%%%%%%%%%%%%%%%%%%%%%%%%%%%%%%%%%%%%%%%%%%%%%%%%%%%%%%%%%%%%%%%%%%%%%%%%

\section*{Light curve analysis}

\begin{figure}[t] \centering \includegraphics[width=\textwidth]{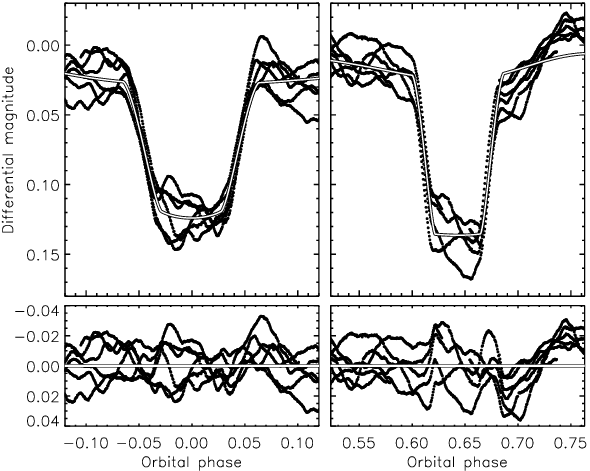} \\
\caption{\label{fig:phase} The TESS light curves of \targ\ from sectors 41, 54 
and 55, with 600~s cadence (filled circles) versus the best fit from {\sc jktebop} 
(white-on-black line) as a function of orbital phase. The primary 
eclipse is shown on the left and the secondary eclipse on the right. The residuals are 
shown on an enlarged scale in the lower panel.} \end{figure}

We modelled the light curve of \targ\ from TESS using version 43 of the {\sc jktebop}\footnote{\texttt{http://www.astro.keele.ac.uk/jkt/codes/jktebop.html}} code \cite{Me++04mn2,Me13aa}. Star~A is formally too deformed to be suitable for {\sc jktebop}, but the intrinsic variability of the system is much more important than the expected bias in the parameters and we were keen to utilise the error estimation algorithms available in the code. We analysed the data with a cadence of 600~s from sectors 41, 54 and 55 simultaneously. Sector 14 was not used due to the lower sampling rate. Conversely, the data with a higher cadence of 120~s available in the last three sectors were not used because it greatly oversamples the changes in brightness due to both eclipses and pulsations.

We fitted for the sum ($r_{\rm A}+r_{\rm B}$) and ratio ($k = {r_{\rm B}}/{r_{\rm A}}$) of the fractional radii of the stars ($r_{\rm A}$ and $r_{\rm B}$), and their central surface brightness ratio in the TESS passband ($J$). We fitted for the orbital period ($P$), reference time of primary minimum ($T_0$), and the eccentricity ($e$) and argument of periastron ($\omega$) in terms of their Poincar\'e elements ($e\cos\omega$ and $e\sin\omega$). A set of straight lines versus time were included for the baseline brightness of the system, one for each half-sector of TESS data, and the coefficients of the lines were included as fitted parameters. We included limb darkening using the simple linear law \cite{Russell12apj2} with the coefficients of both stars fixed to 0.2. More sophisticated laws are not justified due to the strong SLF variability in the light curve, and attempts to fit for the coefficients were unsuccessful for the same reason. We also found that third light was not estimable from the data, so fixed it at zero.

\begin{table} \centering
\caption{\em \label{tab:jktebop} Adopted parameters of \targ\ measured from 
the TESS\ light curves using the {\sc jktebop} code. The uncertainties are 
1$\sigma$ and were determined using residual-permutation simulations.}
\begin{tabular}{lc}
{\em Parameter}                           &       {\em Value}                 \\[3pt]
{\it Fitted parameters:} \\
Time of primary eclipse (BJD$_{\rm TDB}$) & $ 2459438.122    \pm  0.021      $ \\
Orbital period (d)                        & $      13.37441  \pm  0.00075    $ \\
Orbital inclination (\degr)               & $      84.3      \pm  1.0        $ \\
Sum of the fractional radii               & $       0.344    \pm  0.011      $ \\
Ratio of the radii                        & $       0.3006   \pm  0.0090     $ \\
Central surface brightness ratio          & $       1.30     \pm  0.13       $ \\
% Third light                             & $       0.0001   \pm  0.0008     $ \\
% LD coefficient $c$ for star~A           & $       0.548    \pm  0.017      $ \\
% LD coefficient $c$ for star~B           & $       0.516    \pm  0.020      $ \\
% LD coefficient $\alpha$ for star~A      &             0.498 (fixed)          \\
% LD coefficient $\alpha$ for star~B      &             0.467 (fixed)          \\
$e\cos\omega$                             & $       0.2211   \pm  0.0027     $ \\
$e\sin\omega$                             & $      -0.217    \pm  0.021      $ \\
{\it Derived parameters:} \\
Fractional radius of star~A               & $       0.2643   \pm  0.0074     $ \\
Fractional radius of star~B               & $       0.0794   \pm  0.0038     $ \\
Light ratio $\ell_{\rm B}/\ell_{\rm A}$   & $       0.117    \pm  0.010      $ \\[3pt]
Orbital eccentricity                      & $       0.310    \pm  0.014      $ \\
Argument of periastron ($^\circ$)         & $     315.5      \pm  2.9        $ \\
\end{tabular}
\end{table}

The first result of the analysis above is that the secondary eclipse is deeper than the primary (Fig.~\ref{fig:phase}). This conflicts with the standard definition of which is primary and which is secondary, but we have chosen to retain our labelling of the stars so the dominant component remains star~A. From this it can be deduced that star~B has a higher surface brightness, and thus \Teff, than the supergiant star~A. Our results are otherwise very much as expected, and are given in Table\,\ref{tab:jktebop}. The values of $e$ and $\omega$ agree well with previous spectroscopic results.

For the record, we were able to obtain an almost identical fit for the inverse of $k$ (i.e.\ 3.0 versus 0.3). We rejected this solution as being inconsistent with the the model of the system developed by HF84.

%%%%%%%%%%%%%%%%%%%%%%%%%%%%%%%%%%%%%%%%%%%%%%%%%%%%%%%%%%%%%%%%%%%%%%%%%%%%%%%%%%%%%%%%%%%%%%%%%%%%%%%%%%%%%%%%%%%%%%%%%%%%%%%%%%%%%%%%%%%%%%%%%%%%%

\section*{Light curve uncertainties}

The light curve is dominated by the SLF variability, which is essentially red noise from the point of view of eclipse modelling. We therefore used only residual-permutation (RP) simulations \cite{Me08mn} to determine the uncertainties in the fitted parameters. These results are also given in Table\,\ref{tab:jktebop}. Although the data in hand fully cover six orbits of the system, the strong deformation of the eclipses by the SLF signature complicates any attempts to model them. Our errorbars account for this but may still be underestimates.

\begin{figure}[t] \centering \includegraphics[width=\textwidth]{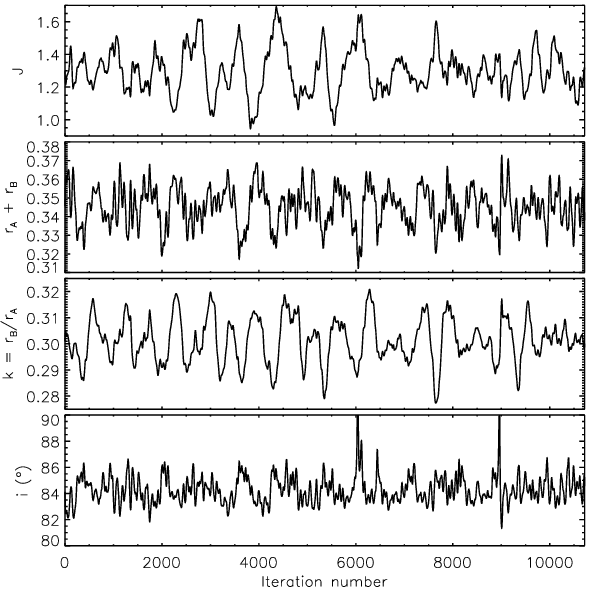} \\
\caption{\label{fig:RPiter} Variation in the best-fitting values of four of the 
photometric parameters during the RP simulations, as the 
residuals are cyclically shifted through the light curve.} \end{figure}

To further illustrate the effect of the SLF variations on the parameters measured from the eclipses, in Fig.~\ref{fig:RPiter} we plot the variation of the best-fitting values of four selected parameters through the RP simulation run. The residuals versus the best {\sc jktebop} fit are shifted by one datapoint between each successive iteration, and the gradual progression of red noise through the light curve causes systematic changes in the fitted parameter values. The most-affected parameter is $J$, which depends primarily on the relative depths of the primary and secondary eclipses. The eclipse depths are significantly changed by the SLF noise, causing a large uncertainty in $J$ and thus the ratios of the \Teff s of the two stars. A similar signature is seen in the ratio of their radii. The variation for $r_{\rm A}+r_{\rm B}$ and $i$ is much faster: these parameters depend on the shapes and durations of the ascending and descending branches of the eclipses, which in turn are shorter than the total eclipse durations. The properties of \targ\ mean it allows a beautiful demonstration of these effects.

\begin{figure}[t] \centering \includegraphics[width=\textwidth]{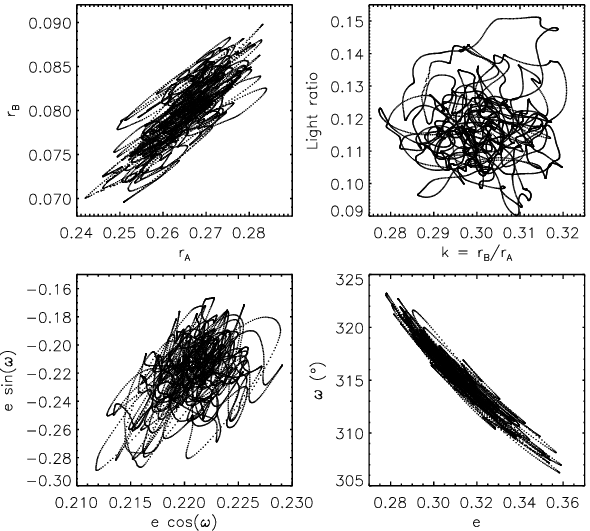} \\
\caption{\label{fig:RPcomp} Comparison plots for the best-fitting values of 
pairs of parameters during the RP simulations.} \end{figure}

Fig.~\ref{fig:RPcomp} shows the variations between pairs of parameters over the RP simulations. The first panel shows $r_{\rm B}$ versus $r_{\rm A}$ and a clear correlation can be seen. The second panel shows the light ratio versus the radius ratio: it has a satisfying child's-scribble appearance but the correlation is small. The inference from this panel is that a spectroscopic light ratio would not be useful in improving the precision of the radius measurements. The remaining two panels show the orbital shape parameters in two forms: the poor determinacy of $e\sin\omega$ (which depends on the ratio of the eclipse durations) is obvious. The much greater correlation between $e$ and $\omega$, versus $e\cos\omega$ and $e\sin\omega$, is also clear: this reiterates the advantage of fitting for the Poincar\'e elements rather than $e$ and $\omega$ directly.

%%%%%%%%%%%%%%%%%%%%%%%%%%%%%%%%%%%%%%%%%%%%%%%%%%%%%%%%%%%%%%%%%%%%%%%%%%%%%%%%%%%%%%%%%%%%%%%%%%%%%%%%%%%%%%%%%%%%%%%%%%%%%%%%%%%%%%%%%%%%%%%%%%%%%

\section*{Stochastic low-frequency variability}

\begin{figure}[t] \centering \includegraphics[width=\textwidth]{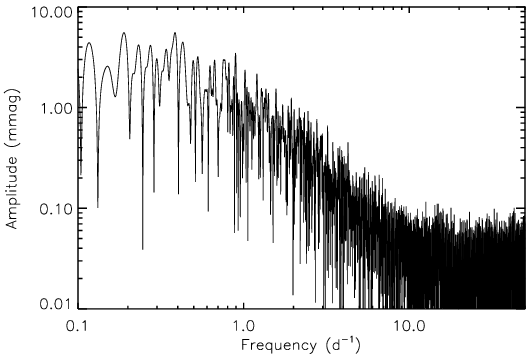} \\
\caption{\label{fig:fourier} Periodogram of the residuals of the best fit for 
TESS sectors 54 and 55, calculated using the {\sc period04} code.} \end{figure}

The intrinsic variability in the light curves of \targ\ is obvious and can safely be assumed to arise from the supergiant star~A. To investigate it further we used the best fit from {\sc jktebop} found above, restricted it to sectors 54 and 55 as these data are semi-continuous, selected the residuals of the best fit, and calculated a periodogram using the {\sc period04} code \cite{LenzBreger04iaus}. The result is shown in Fig.~\ref{fig:fourier}. A periodogram of the 120~s cadence data from sector 55 shows no significant signal at higher frequencies, up to the Nyquist limit for these data of 350~d$^{-1}$.

Fig.~\ref{fig:fourier} shows excess power at frequencies below 5~d$^{-1}$ with a large number of peaks with significant amplitude. This is characteristic of SLF variability \cite{Bowman+19aa,Bowman+19natas}, has been seen before in dEBs \cite{Tkachenko+14mn,MeBowman22mn}, and is attributable to internal gravity waves excited at the boundary of a convective region within the star \cite{Rogers+13apj,AertsRogers15apj}. The two highest peaks occur at 0.19 and 0.38~d$^{-1}$ and have amplitudes of 5.6~mmag.

%%%%%%%%%%%%%%%%%%%%%%%%%%%%%%%%%%%%%%%%%%%%%%%%%%%%%%%%%%%%%%%%%%%%%%%%%%%%%%%%%%%%%%%%%%%%%%%%%%%%%%%%%%%%%%%%%%%%%%%%%%%%%%%%%%%%%%%%%%%%%%%%%%%%%

\section*{Spectroscopic properties}

\begin{sidewaysfigure} \centering
\includegraphics[width=\textwidth]{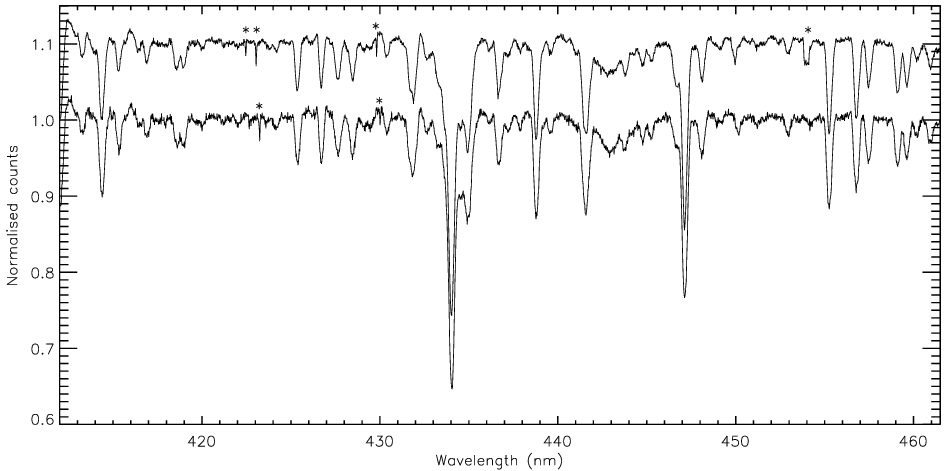}
\caption{\label{fig:spec} The two combined spectra of \targ\ taken when the RVs of the 
two stars were the same (normalised to 1.0) and when they show their greatest separation 
(offset by +0.1). The second spectrum has been shifted by $-0.2$~nm to align the spectral 
lines of star~A. CCD cosmetics are indicated with asterisks.} \end{sidewaysfigure}

\targ\ is spectroscopically difficult due to the large line broadening and SLF-induced line profile variability of star~A. HF84 identified faint peaks in the cross-correlation functions arising from star~B which was found to be approximately ten times fainter than star~A at blue-optical wavelengths. This was questioned by Popper \cite{Popper93pasp}, who was not able to confidently identify lines of star~B despite having spectra of much better quality. A relevant point here is that the light ratio we found from the TESS light curve matches that inferred by HF84 from the relative areas of the cross-correlation function peaks.

As described above, we obtained two INT/IDS spectra in order to investigate this further. These are shown in Fig.~\ref{fig:spec}, where the spectrum from the second night has been offset by +0.1 from that for the first night, and also shifted by $-0.20$~nm to remove the RV variation of star~A relative to the first spectrum. The first spectrum was taken at orbital phase 0.617 -- at the beginning of totality during secondary eclipse -- so contains light from star~A only. The second spectrum was taken at phase 0.768 so includes light from both stars, with a velocity separation of 376\kms.

In both cases the spectra are corrected for the barycentric velocity. The spectra show strong H, He and O absorption, plus C, N, Si and Mg lines and a diffuse interstellar band centred at approximately 443~nm. The strong O~{\sc ii} lines at 4348 and 4416~\AA\ confirm the supergiant classification, and are strong enough to support a luminosity class of 1a rather than the typically-quoted 1b.

\begin{figure}[t] \centering \includegraphics[width=\textwidth]{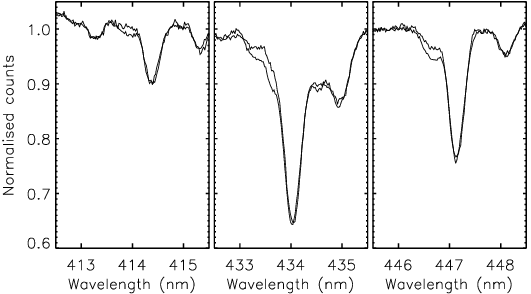} \\
\caption{\label{fig:lines} Comparison between the two combined spectra of \targ\ in the 
region of the He~{\sc i} 4143~\AA, H$\gamma$ 4340~\AA\ and He~{\sc i} 4471~\AA\ lines. The two spectra 
were aligned in wavelength to make the strong lines from star~A overlap.} \end{figure}

A careful comparison of the two spectra reveals the appearance of three faint absorption lines to the left of the main lines on the second night (see Fig.~\ref{fig:lines}). These are exactly where we would expect to find lines from star~B, and similar features are \emph{not} seen where they should not be (e.g.\ for the O~{\sc ii} lines). This suggests that \targ\ may well be double-lined and thus suitable for direct measurement of its masses and radii. This work will be helped by obtaining extensive new high-quality spectroscopy and analysing them using methods not available to previous workers such as two-dimensional cross-correlation \cite{ZuckerMazeh94apj}, broadening functions \cite{Rucinski99aspc} and spectral disentangling \cite{SimonSturm94aa}. The last method is most promising (e.g.\ ref.~\cite{Pavlovski+09mn}) but may be affected by the line-profile variations from the stochastic variability. We defer further analysis until suitable spectra are available.

% #     TIME       RV_STAR_A    ERROR      PHASE        MODEL      (O-C)
%   60128.483620     0.00000 999.000000   0.61740217   -3.75019    3.75019
%   60130.493330     0.00000 999.000000   0.76766473  125.46213 -125.46213
% #     TIME       RV_STAR_B    ERROR      PHASE        MODEL      (O-C)
%   60128.483620     0.00000 999.000000   0.61740217    7.50039   -7.50039
%   60130.493330     0.00000 999.000000   0.76766473 -250.92425  250.92425

%%%%%%%%%%%%%%%%%%%%%%%%%%%%%%%%%%%%%%%%%%%%%%%%%%%%%%%%%%%%%%%%%%%%%%%%%%%%%%%%%%%%%%%%%%%%%%%%%%%%%%%%%%%%%%%%%%%%%%%%%%%%%%%%%%%%%%%%%%%%%%%%%%%%%%

\section*{Physical properties of \targ}

This work presents the first determinate solution of the light curve of \targ\ and thus enables a more direct estimation of the properties of the system. For this the velocity amplitudes of the stars' spectroscopic orbits are needed. There is only one source available in the literature, HF84, and their results were questioned by Popper \cite{Popper93pasp}. We chose to adopt (approximately) the values from HF84 but with increased errorbars to account for the conflicting results: $K_{\rm A} = 103 \pm 2$\kms\ and $K_{\rm B} = 206 \pm 10$\kms.

HF84 adopted a \Teff\ of 25\,000~K for star~A from a calibration versus spectral type by Underhill et al.\ \cite{Underhill+79mn}, to which we add a plausible errorbar of 2000~K. The surface brightness ratio from the light curve analysis (Table~\ref{tab:jktebop}) then gives a \Teff\ of $26500 \pm 2500$~K for star~B, which implies a spectral type of B1~V using the calibration of Pecaut \& Mamajek \cite{PecautMamajek13apjs}. Wu et al.\ \cite{Wu+11aa} gave a higher \Teff\ of $26556 \pm 1934$~K for the system (analysed as if it were a single star) but we did not use this value as it yielded a distance to the system significantly longer than that from the \gaia\ parallax (see below). More precise and accurate \Teff\ values could be obtained from spectroscopy of the system in future.

\begin{table} \centering
\caption{\em Plausible physical properties of \targ\ defined using the nominal solar 
units given by IAU 2015 Resolution B3 (ref.\ \cite{Prsa+16aj}). \label{tab:absdim}}
\begin{tabular}{lr@{\,$\pm$\,}lr@{\,$\pm$\,}l}
{\em Parameter}        & \multicolumn{2}{c}{\em Star A} & \multicolumn{2}{c}{\em Star B}    \\[3pt]
Mass ratio   $M_{\rm B}/M_{\rm A}$          & \multicolumn{4}{c}{$0.50 \pm 0.02$}           \\
Semimajor axis of relative orbit (\Rsunnom) & \multicolumn{4}{c}{$78 \pm 2$}                \\
Mass (\Msunnom)                             & 23      & 2           & 11.9    & 0.7         \\
Radius (\Rsunnom)                           & 20.6    & 0.8         &  6.2    & 0.3         \\
Surface gravity ($\log$[cgs])               &  3.19   & 0.03        &  3.93   & 0.04        \\
% Density ($\!\!$\rhosun)                   &  0.398  & 0.027       &  0.550  & 0.039       \\
% Synchronous rotational velocity ($\!\!$\kms)&0.93   & 0.92        & 35.89   & 0.85        \\
Effective temperature (K)                   &  25000  & 2000        &  26500  & 2500        \\
Luminosity $\log(L/\Lsunnom)$               &   5.18  & 0.14        &   4.23  & 0.17        \\
$M_{\rm bol}$ (mag)                         &$-$8.2   & 0.4         &$-$5.8   & 0.4         \\
Distance (pc)                               & \multicolumn{4}{c}{$1520 \pm 110$}            \\[3pt]
\end{tabular}
\end{table}

Armed with these numbers, we calculated the expected physical properties of \targ\ using using the {\sc jktabsdim} code \cite{Me++05aa} in our usual way for this series of papers. However, in this case, the numbers should not be taken as definitive due to the disagreement over whether the RVs of star~B are reliable. The inferred properties are given in Table~\ref{tab:absdim}. 

To determine the distance to the system we used the Tycho-2 $B$ and $V$ magnitudes \cite{Hog+00aa} which are averages of 14 measurements each, the 2MASS $JHK_s$ magnitudes \cite{Cutri+03book} which are single measurements taken at orbital phase 0.78 (i.e.\ outside eclipse), and bolometric corrections from Girardi et al.\ \cite{Girardi+02aa}. An interstellar reddening of $\EBV = 0.43 \pm 0.10$~mag is needed to bring the $BV$ and $JHK_s$ distances into agreement, giving a final $K$-band distance of $1520 \pm 110$~pc which is concordant with the distance of $1450 \pm 53$~pc from the \gaia\ DR3 parallax. This agreement supports the reliability of the approximate system parameters put forward in Table~\ref{tab:absdim}.

%%%%%%%%%%%%%%%%%%%%%%%%%%%%%%%%%%%%%%%%%%%%%%%%%%%%%%%%%%%%%%%%%%%%%%%%%%%%%%%%%%%%%%%%%%%%%%%%%%%%%%%%%%%%%%%%%%%%%%%%%%%%%%%%%%%%%%%%%%%%%%%%%%%%%%

\section*{The similarity of \targ\ and V380~Cyg}

One object stands out as being rather similar to \targ. V380~Cyg is a dEB containing B1~III and B2~V components with an orbital period of 12.4~d, an eccentricity of 0.222, and an extensive observational history \cite{HillBatten84aa,PopperGuinan98pasp,Guinan+00apj,Pavlovski+09mn,Tkachenko+12mn,Tkachenko+14mn}.

\begin{figure}[t] \centering \includegraphics[width=\textwidth]{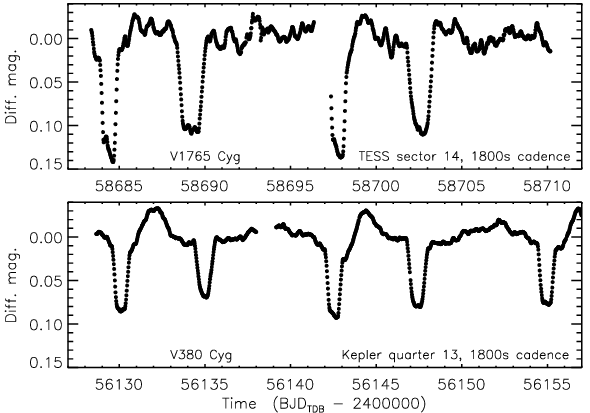} \\
\caption{\label{fig:comp} Comparison between the TESS light curve of \targ\ (top) 
and the \kepler\ light curve of V380~Cyg (bottom). The y-axes are the same, and 
the x-axes are of the same duration, in the two panels.} \end{figure}

In Fig.~\ref{fig:comp} we show light curves of the two systems with the same axis scales. We chose TESS sector 14 for \targ\ and \kepler\ \cite{Borucki16rpph} quarter 13 for V380~Cyg, in both cases with a sampling rate of 1800~s. It can be seen that the eclipses are slightly deeper and longer in \targ, and in particular the SLF variability is much stronger. Although the more evolved components in the two systems have almost the same fractional radii, V380~Cyg has a much more pronounced ``orbital hump'' at periastron passage shortly after primary eclipse.

Tkachenko et al. \cite{Tkachenko+14mn} found these masses and radii for the components of V380~Cyg: $M_{\rm A} = 11.43 \pm 0.19$\Msun, $R_{\rm A} = 15.71 \pm 0.13$\Rsun, $M_{\rm B} = 7.00 \pm 0.14$\Msun\ and $R_{\rm B} = 3.82 \pm 0.05$\Rsun. \targ\ is therefore a more extreme version of V380~Cyg. Tkachenko et al.\ used 406 spectra in their investigation -- \targ\ would likely need a similar amount because the larger light ratio (so star~B is relatively brighter) will be offset by the stronger variability of star~A. Whilst this is a \emph{lot} of spectra, the brightness of the system means such a number is achievable.

%%%%%%%%%%%%%%%%%%%%%%%%%%%%%%%%%%%%%%%%%%%%%%%%%%%%%%%%%%%%%%%%%%%%%%%%%%%%%%%%%%%%%%%%%%%%%%%%%%%%%%%%%%%%%%%%%%%%%%%%%%%%%%%%%%%%%%%%%%%%%%%%%%%%%%

\section*{Comparison with theoretical models}

\begin{figure}[t] \centering \includegraphics[width=\textwidth]{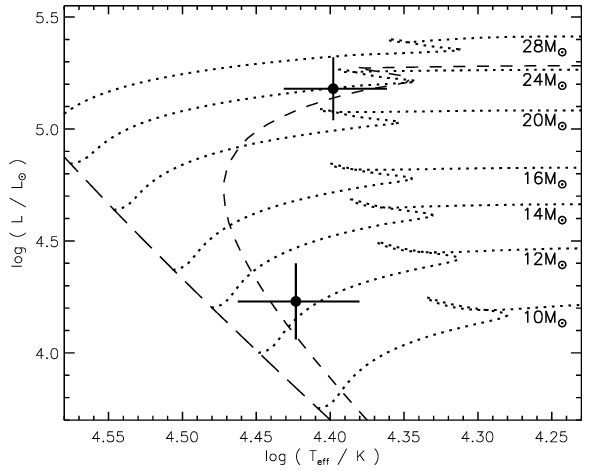} \\
\caption{\label{fig:hrd} Hertzsprung-Russell diagram for the components of 
\targ\ (filled circles with errorbars) and the predictions of the {\sc parsec} 
1.2S models \cite{Chen+14mn} for selected masses (dotted lines with masses 
labelled). The zero-age main sequence is indicated with a long-dashed line, 
and a 7-Myr isochrone with a short-dashed line.} \end{figure}

Although the properties in Table~\ref{tab:absdim} are not definitive, a brief check against theoretical predictions could be illuminating. For this we adopted the {\sc parsec} 1.2S models from Chen et al.\ \cite{Chen+14mn}. A reasonable agreement is found in plots of mass versus radius and \Teff\ (not shown) for a metal abundance of $Z=0.02$ and an age of $7 \pm 1$~Myr. This supports the lower of the two \Teff\ measurements discussed above. The radius of star~B is approximately 3$\sigma$ larger than predicted, but the two \Teff\ values sit perfectly on the predictions.

Fig.~\ref{fig:hrd} shows a Hertzsprung-Russell diagram with the components of \targ\ and predictions from the {\sc parsec} models for a range of masses. The figure includes the zero-age main sequence and an isochrone for an age of 7~Myr. The agreement between observation and theory is good. More precise properties of \targ\ are needed to provide a useful test of the models.

%%%%%%%%%%%%%%%%%%%%%%%%%%%%%%%%%%%%%%%%%%%%%%%%%%%%%%%%%%%%%%%%%%%%%%%%%%%%%%%%%%%%%%%%%%%%%%%%%%%%%%%%%%%%%%%%%%%%%%%%%%%%%%%%%%%%%%%%%%%%%%%%%%%%%%

\section*{Summary and conclusions}

\targ\ is a very interesting totally-eclipsing binary containing a B0.5 supergiant and a B1 main-sequence star on a 13.37~d orbit with an eccentricity of 0.315. Extensive previous work has yielded a reliable spectroscopic orbit for star~A and a less reliable one for star~B. Whilst the reality of the detection of star~B in the spectra has been questioned, the resulting RVs lead us to a plausible set of properties for the system. 

In this work we analysed four sectors of observations from the TESS mission, allowing us to determine the radii of the stars from the eclipse profiles. Previous radius estimates were based only on calibrations versus spectral type. We arrive at physical properties in agreement with published values but on a more solid empirical basis. These properties can be matched by the {\sc parsec} models for a solar chemical composition and an age in the region of 7~Myr. 

Star~A shows strong stochastic brightness variations of the SLF type, which distort the eclipse shapes and complicate both photometric and spectroscopic analyses. More extensive photometry, should the opportunity arise, may allow specific pulsation modes to be identified. \targ\ is similar to but a more extreme version of the well-studied V380~Cyg system.

Apsidal motion has been detected in this system \cite{Raja94aa,Raja94aa}, although the detection has been questioned (HF84). The value of $\omega$ we deduced from the light curve supports the existence of apsidal motion, as it is significantly greater than the values found in the old spectroscopic studies. A more detailed analysis of this phenomenon would be rewarding.

We also presented two epochs of medium-resolution spectroscopy which confirm the spectral classification of star~A. Star~B produces approximately 10\% of the light of the system and our spectra show evidence of its absorption lines which encourages further study. We strongly recommend that a large set of high-quality spectra are obtained for this system to confirm the detection of star~B and for measurement of the atmospheric parameters and spectroscopic orbits of both stars. Such work will be difficult, but will be helped by the development of new analysis tools since the last detailed spectroscopic study of this system. 

Finally, we note that TESS is scheduled to observe \targ\ in four further sectors (74, 75, 81 and 82) during 2024. The addition of these data to the analysis should significantly improve the measurements of the radii of the stars, which are currently limited by the pulsations interfering with the eclipse shapes. The nature of the \targ\ system makes such work well worth pursuing.

%%%%%%%%%%%%%%%%%%%%%%%%%%%%%%%%%%%%%%%%%%%%%%%%%%%%%%%%%%%%%%%%%%%%%%%%%%%%%%%%%%%%%%%%%%%%%%%%%%%%%%%%%%%%%%%%%%%%%%%%%%%%%%%%%%%%%%%%%%%%%%%%%%%%%%

\section*{Acknowledgements}

I am grateful to Steve Overall for assistance in obtaining the INT spectra, where by ``assistance'' I mean he actually performed the observations whilst I sat there, watched, and drank tea.
% and an anonymous referee for useful comments on a draft of this work.
%We thank the anonymous referee for a quick and positive report.
This paper includes data collected by the TESS\ mission and obtained from the MAST data archive at the Space Telescope Science Institute (STScI). Funding for the TESS\ mission is provided by the NASA's Science Mission Directorate. STScI is operated by the Association of Universities for Research in Astronomy, Inc., under NASA contract NAS 5–26555.
%This work has made use of data from the European Space Agency (ESA) mission {\it Gaia}\footnote{\texttt{https://www.cosmos.esa.int/gaia}}, processed by the {\it Gaia} Data Processing and Analysis Consortium (DPAC\footnote{\texttt{https://www.cosmos.esa.int/web/gaia/dpac/consortium}}). Funding for the DPAC has been provided by national institutions, in particular the institutions participating in the {\it Gaia} Multilateral Agreement.
The following resources were used in the course of this work: the NASA Astrophysics Data System; the SIMBAD database operated at CDS, Strasbourg, France; and the ar$\chi$iv scientific paper preprint service operated by Cornell University.

%%%%%%%%%%%%%%%%%%%%%%%%%%%%%%%%%%%%%%%%%%%%%%%%%%%%%%%%%%%%%%%%%%%%%%%%%%%%%%%%%%%%%%%%%%%%%%%%%%%%%%%%%%%%%%%%%%%%%%%%%%%%%%%%%%%%%%%%%%%%%%%%%%%%%

% \bibliographystyle{obsmaga}
% \bibliography{jkt}

\begin{thebibliography}{10}
\newcommand{\enquote}[1]{`(#1)'}

\bibitem{Andersen91aarv}
J.~{Andersen}, \textit{A\&ARv}, \textbf{3}, 91, 1991.

\bibitem{Torres++10aarv}
G.~{Torres}, J.~{Andersen} \& A.~{Gim{\'e}nez}, \textit{A\&ARv}, \textbf{18},
  67, 2010.

\bibitem{Me15aspc}
J.~{Southworth}, in \textit{Living Together: Planets, Host Stars and Binaries}
  (S.~M. {Rucinski}, G.~{Torres} \& M.~{Zejda}, eds.), 2015,
  \textit{Astronomical Society of the Pacific Conference Series}, vol. 496, p.
  321.

\bibitem{Hoxie73aa}
D.~T. {Hoxie}, \textit{A\&A}, \textbf{26}, 437, 1973.

\bibitem{Lacy77apjs}
C.~H. {Lacy}, \textit{ApJS}, \textbf{34}, 479, 1977.

\bibitem{Torres13an}
G.~{Torres}, \textit{Astronomische Nachrichten}, \textbf{334}, 4, 2013.

\bibitem{Herrero+92aa}
A.~{Herrero} \textit{et~al.}, \textit{A\&A}, \textbf{261}, 209, 1992.

\bibitem{Tkachenko+20aa}
A.~{Tkachenko} \textit{et~al.}, \textit{A\&A}, \textbf{637}, A60, 2020.

\bibitem{Sana+14apjs}
H.~{Sana} \textit{et~al.}, \textit{ApJS}, \textbf{215}, 15, 2014.

\bibitem{Kobulnicky+14apjs}
H.~A. {Kobulnicky} \textit{et~al.}, \textit{ApJS}, \textbf{213}, 34, 2014.

\bibitem{Bowman+19aa}
D.~M. {Bowman} \textit{et~al.}, \textit{A\&A}, \textbf{621}, A135, 2019.

\bibitem{MeBowman22mn}
J.~{Southworth} \& D.~M. {Bowman}, \textit{MNRAS}, \textbf{513}, 3191, 2022.

\bibitem{Me+20mn}
J.~{Southworth} \textit{et~al.}, \textit{MNRAS}, \textbf{497}, L19, 2020.

\bibitem{LeeHong21aj}
J.~W. {Lee} \& K.~{Hong}, \textit{AJ}, \textbf{161}, 32, 2021.

\bibitem{Me++21mn}
J.~{Southworth}, D.~M. {Bowman} \& K.~{Pavlovski}, \textit{MNRAS},
  \textbf{501}, L65, 2021.

\bibitem{Kahraman+17mn}
F.~{Kahraman Ali{\c{c}}avu{\c{s}}} \textit{et~al.}, \textit{MNRAS},
  \textbf{470}, 915, 2017.

\bibitem{Chen+22apjs}
X.~{Chen} \textit{et~al.}, \textit{ApJS}, \textbf{263}, 34, 2022.

\bibitem{Me21obs6}
J.~{Southworth}, \textit{The Observatory}, \textbf{141}, 282, 2021.

\bibitem{Me++23mn}
J.~{Southworth}, S.~J. {Murphy} \& K.~{Pavlovski}, \textit{MNRAS},
  \textbf{520}, L53, 2023.

\bibitem{Debosscher+13aa}
J.~{Debosscher} \textit{et~al.}, \textit{A\&A}, \textbf{556}, A56, 2013.

\bibitem{Lee16apj}
J.~W. {Lee}, \textit{ApJ}, \textbf{833}, 170, 2016.

\bibitem{MeVanreeth22mn}
J.~{Southworth} \& T.~{Van Reeth}, \textit{MNRAS}, \textbf{515}, 2755, 2022.

\bibitem{Kurtz+20mn}
D.~W. {Kurtz} \textit{et~al.}, \textit{MNRAS}, \textbf{494}, 5118, 2020.

\bibitem{Handler+20natas}
G.~{Handler} \textit{et~al.}, \textit{Nature Astronomy}, \textbf{4}, 684, 2020.

\bibitem{Me20obs}
J.~{Southworth}, \textit{The Observatory}, \textbf{140}, 247, 2020.

\bibitem{Me21univ}
J.~{Southworth}, \textit{Universe}, \textbf{7}, 369, 2021.

\bibitem{Gaia21aa}
{Gaia Collaboration}, \textit{A\&A}, \textbf{649}, A1, 2021.

\bibitem{HoffleitJaschek91}
D.~{Hoffleit} \& C.~. {Jaschek}, \textit{{The Bright Star Catalogue}} (New
  Haven, Conn.: Yale University Observatory, 1991, 5th ed.), 1991.

\bibitem{CannonPickering23anhar}
A.~J. {Cannon} \& E.~C. {Pickering}, \textit{Annals of Harvard College
  Observatory}, \textbf{98}, 1, 1923.

\bibitem{Stassun+19aj}
K.~G. {Stassun} \textit{et~al.}, \textit{AJ}, \textbf{158}, 138, 2019.

\bibitem{Hog+00aa}
E.~{H{\o}g} \textit{et~al.}, \textit{A\&A}, \textbf{355}, L27, 2000.

\bibitem{Cutri+03book}
R.~M. {Cutri} \textit{et~al.}, \textit{{2MASS All Sky Catalogue of Point
  Sources}} (The IRSA 2MASS All-Sky Point Source Catalogue, NASA/IPAC Infrared
  Science Archive, Caltech, US), 2003.

\bibitem{MorganRoman50apj}
W.~W. {Morgan} \& N.~G. {Roman}, \textit{ApJ}, \textbf{112}, 362, 1950.

\bibitem{PlaskettPearce31pdao}
J.~S. {Plaskett} \& J.~A. {Pearce}, \textit{PDAO}, \textbf{5}, 1, 1931.

\bibitem{MayerChochol81pasp}
P.~{Mayer} \& D.~{Chochol}, \textit{PASP}, \textbf{93}, 608, 1981.

\bibitem{PercyWelch83pasp}
J.~R. {Percy} \& D.~L. {Welch}, \textit{PASP}, \textbf{95}, 491, 1983.

\bibitem{Morgan++55apjs}
W.~W. {Morgan}, A.~D. {Code} \& A.~E. {Whitford}, \textit{ApJS}, \textbf{2},
  41, 1955.

\bibitem{Hiltner56apjs}
W.~A. {Hiltner}, \textit{ApJS}, \textbf{2}, 389, 1956.

\bibitem{Hiltner51apj}
W.~A. {Hiltner}, \textit{ApJ}, \textbf{114}, 241, 1951.

\bibitem{Lesh68apjs}
J.~R. {Lesh}, \textit{ApJS}, \textbf{17}, 371, 1968.

\bibitem{HillFisher84aa}
G.~{Hill} \& W.~A. {Fisher}, \textit{A\&A}, \textbf{139}, 123, 1984.

\bibitem{Mayer+91baicz2}
P.~{Mayer} \textit{et~al.}, \textit{BAICz}, \textbf{42}, 230, 1991.

\bibitem{Raja94aa}
T.~{Raja}, \textit{A\&A}, \textbf{284}, 82, 1994.

\bibitem{Popper93pasp}
D.~M. {Popper}, \textit{PASP}, \textbf{105}, 721, 1993.

\bibitem{PercyKhaja95jrasc}
J.~R. {Percy} \& N.~{Khaja}, \textit{JRASC}, \textbf{89}, 91, 1995.

\bibitem{Talens+17aa}
G.~J.~J. {Talens} \textit{et~al.}, \textit{A\&A}, \textbf{601}, A11, 2017.

\bibitem{Burggraaff+18aa}
O.~{Burggraaff} \textit{et~al.}, \textit{A\&A}, \textbf{617}, A32, 2018.

\bibitem{Ricker+15jatis}
G.~R. {Ricker} \textit{et~al.}, \textit{Journal of Astronomical Telescopes,
  Instruments, and Systems}, \textbf{1}, 014003, 2015.

\bibitem{Lightkurve18}
{Lightkurve Collaboration}, \enquote{{Lightkurve: Kepler and TESS time series
  analysis in Python}}, Astrophysics Source Code Library, 2018.

\bibitem{Jenkins+16spie}
J.~M. {Jenkins} \textit{et~al.}, in \textit{Proc.\ SPIE}, 2016, \textit{Society
  of Photo-Optical Instrumentation Engineers (SPIE) Conference Series}, vol.
  9913, p. 99133E.

\bibitem{Me22obs6}
J.~{Southworth}, \textit{The Observatory}, \textbf{142}, 267, 2022.

\bibitem{Me++04mn2}
J.~{Southworth}, P.~F.~L. {Maxted} \& B.~{Smalley}, \textit{MNRAS},
  \textbf{351}, 1277, 2004.

\bibitem{Me13aa}
J.~{Southworth}, \textit{A\&A}, \textbf{557}, A119, 2013.

\bibitem{Russell12apj2}
H.~N. {Russell}, \textit{ApJ}, \textbf{36}, 54, 1912.

\bibitem{Me08mn}
J.~{Southworth}, \textit{MNRAS}, \textbf{386}, 1644, 2008.

\bibitem{LenzBreger04iaus}
P.~{Lenz} \& M.~{Breger}, in \textit{The A-Star Puzzle, Cambridge University
  Press, Cambridge, UK.} ({J.~Zverko, J.~{\v{Z}}i{\v{z}}novsky, S.~J.~Adelman,
  \& W.~W.~Weiss}, ed.), 2004, \textit{IAU Symposium}, vol. 224, pp. 786--790.

\bibitem{Bowman+19natas}
D.~M. {Bowman} \textit{et~al.}, \textit{Nature Astronomy}, \textbf{3}, 760,
  2019.

\bibitem{Tkachenko+14mn}
A.~{Tkachenko} \textit{et~al.}, \textit{MNRAS}, \textbf{438}, 3093, 2014.

\bibitem{Rogers+13apj}
T.~M. {Rogers} \textit{et~al.}, \textit{ApJ}, \textbf{772}, 21, 2013.

\bibitem{AertsRogers15apj}
C.~{Aerts} \& T.~M. {Rogers}, \textit{ApJ}, \textbf{806}, L33, 2015.

\bibitem{ZuckerMazeh94apj}
S.~{Zucker} \& T.~{Mazeh}, \textit{ApJ}, \textbf{420}, 806, 1994.

\bibitem{Rucinski99aspc}
S.~{Rucinski}, in \textit{IAU Colloq.\ 170: Precise Stellar Radial Velocities}
  ({J.~B.~Hearnshaw \& C.~D.~Scarfe}, ed.), 1999, \textit{Astronomical Society
  of the Pacific Conference Series}, vol. 185, p.~82.

\bibitem{SimonSturm94aa}
K.~P. {Simon} \& E.~{Sturm}, \textit{A\&A}, \textbf{281}, 286, 1994.

\bibitem{Pavlovski+09mn}
K.~{Pavlovski} \textit{et~al.}, \textit{MNRAS}, \textbf{400}, 791, 2009.

\bibitem{Underhill+79mn}
A.~B. {Underhill} \textit{et~al.}, \textit{MNRAS}, \textbf{189}, 601, 1979.

\bibitem{PecautMamajek13apjs}
M.~J. {Pecaut} \& E.~E. {Mamajek}, \textit{ApJS}, \textbf{208}, 9, 2013.

\bibitem{Wu+11aa}
Y.~{Wu} \textit{et~al.}, \textit{A\&A}, \textbf{525}, A71, 2011.

\bibitem{Prsa+16aj}
A.~{Pr{\v s}a} \textit{et~al.}, \textit{AJ}, \textbf{152}, 41, 2016.

\bibitem{Me++05aa}
J.~{Southworth}, P.~F.~L. {Maxted} \& B.~{Smalley}, \textit{A\&A},
  \textbf{429}, 645, 2005.

\bibitem{Girardi+02aa}
L.~{Girardi} \textit{et~al.}, \textit{A\&A}, \textbf{391}, 195, 2002.

\bibitem{HillBatten84aa}
G.~{Hill} \& A.~H. {Batten}, \textit{A\&A}, \textbf{141}, 39, 1984.

\bibitem{PopperGuinan98pasp}
D.~M. {Popper} \& E.~F. {Guinan}, \textit{PASP}, \textbf{110}, 572, 1998.

\bibitem{Guinan+00apj}
E.~F. {Guinan} \textit{et~al.}, \textit{ApJ}, \textbf{544}, 409, 2000.

\bibitem{Tkachenko+12mn}
A.~{Tkachenko} \textit{et~al.}, \textit{MNRAS}, \textbf{424}, L21, 2012.

\bibitem{Borucki16rpph}
W.~J. {Borucki}, \textit{Reports on Progress in Physics}, \textbf{79}, 036901,
  2016.

\bibitem{Chen+14mn}
Y.~{Chen} \textit{et~al.}, \textit{MNRAS}, \textbf{444}, 2525, 2014.

\end{thebibliography}

%%%%%%%%%%%%%%%%%%%%%%%%%%%%%%%%%%%%%%%%%%%%%%%%%%%%%%%%%%%%%%%%%%%%%%%%%%%%%%%%%%%%%%%%%%%%%%%%%%%%%%%%%%%%%%%%%%%%%%%%%%%%%%%%%%%%%%%%%%%%%%%%%%%%%
\end{document}